\documentclass[pre,aps,showpacs,nofootinbib]{revtex4}
\usepackage{graphicx}
\usepackage{natbib}
\usepackage{bm}
\usepackage{amssymb}
\usepackage{amsmath}
\usepackage{euscript}
\usepackage{esint}
\usepackage{prelim2e}
\usepackage{slashed}

\begin{document}

\title{An interpretation of the infrared singularity of the effective electromagnetic field}

\author{Kirill A. Kazakov}
\author{Vladimir V. Nikitin}
\affiliation{%
Department of Theoretical Physics, Physics Faculty, Moscow State
University, 119991, Moscow, Russian Federation}

\begin{abstract}
The problem of infrared divergence of the effective electromagnetic field produced by elementary particles is revisited using the non-equilibrium model of an electron interacting with low-temperature photons. It is argued that the infrared singularity of the effective field can be interpreted as a thermalization of the electron. It is shown that this thermalization is negligible in actual field measurements as it is completely dominated by the usual quantum spreading.

\end{abstract}
 \maketitle

\subsection{Introduction}

It is well known that calculations in quantum field theories involving massless particles are plagued by the presence of infrared divergences. Unlike the case of ultraviolet singularities which can in general be consistently ``subtracted'' by an appropriate redefinition of the parameters of the theory, there is no unique recipe to deal with/interpret the infrared-infinite contributions. As far as the scattering matrix method is applicable, a general prescription is to sum the scattering cross sections over suitable classes of initial and/or final states. In quantum electrodynamics or quantum gravity, for instance, the standard resolution of the ``infrared catastrophe'' is to sum over all final states containing arbitrary number of real soft photons or gravitons \cite{bloch} (in Yang-Mills theories, the procedure is more intricate \cite{leenaum}). However, the problem persists beyond the scope of the S-matrix, in particular, it presents a serious obstacle to the use of the effective field methods which are of vital importance in investigating such issues as the spontaneous symmetry breaking in the electroweak theory, the quark-gluon plasma effects in quantum chromodynamics, particle creation and inflation in quantum cosmology, {\it etc.}

Specifically, it is known that the vertex formfactors of the massive field quanta, used in constructing the effective (mean) fields of elementary particles, are  infrared-divergent. Though the same formfactors appear in the scattering amplitudes, the above-mentioned prescription of the S-matrix theory does not apply to the effective fields. Indeed, the very statement of the problem in the two cases is quite different. The Bloch-Nordsieck theorem states that infrared divergences in the radiative corrections to the scattering cross sections exactly cancel those due to the emission of real soft photons. At the same time, real photons do not appear at all in the calculation of the static effective field (at zero temperature). By this reason, the effective field formalism can be applied, strictly speaking, only to the fields produced by classical sources. If, for instance, the charged particle mass is sufficiently large, radiative corrections to its interaction with the electromagnetic field can be neglected, thus putting the question about their divergence aside. In fact, assumptions of this kind underlie the classic calculation \cite{serber} of the effective static field of a point source and its generalizations \cite{bialb}. However, this limitation looks rather unsatisfactory from the theoretical point of view, especially in the light of measurability of the electromagnetic field, established long ago \cite{measurability}.

This problem is sharpened at finite temperatures, because the heat-bath effects introduce new low-energy singularities to the photon propagator. These additional singularities are generally believed to worsen infrared properties of the Feynman integrals. At the same time, inclusion of the finite temperature effects into consideration is a question of principle: no matter how small temperature is, it is never zero exactly, and the question whether $T\approx 0$ can be replaced by $T=0$ can be answered only by investigating the general nonzero-temperature case. In other words, continuity of this sort, being a necessary physical requirement for the very possibility to neglect the heat-bath effects, is to be proved rather than postulated. Previous investigations of the infrared problem have been aimed mainly at generalizing the Bloch-Nordsieck and Kinoshita-Lee-Nauenberg theorems to nonzero temperatures \cite{tempsmatrix}. As was already mentioned, their results do not apply to the effective fields.

The aim of this Letter is to propose a physical interpretation of the infrared divergence of the effective electromagnetic field produced by an elementary charged particle. Namely, we shall evaluate the effective Coulomb field of a free electron that was in an arbitrary state in the past, and show that this field vanishes upon account of the infrared radiative corrections to all orders of the perturbation theory. We will argue that this result signifies existence of a peculiar spreading of the charge which is inherently irreversible, in the sense that it makes formally impossible preparation of a spatially localized free particle state at finite times by operating with arbitrary states in the remote past.

Our consideration applies equally to the zero- and nonzero-temperature cases, and begins in the next section with a description of the physical model to be investigated. We then introduce an infrared regularization of the model, which represents a modification of the usual momentum cutoff method, and prove that the proposed scheme admits factorization of the infrared radiative contributions to the effective field to all orders of the perturbation theory. This result is used in the last section to demonstrate that the Coulomb field of the electron vanishes at any given position in the limit of removed cutoff, in a way that respects the total charge conservation. We argue that interpreted in terms of the electron density matrix, this field nullification can be described as an electron thermalization through its interaction with photons.

\subsection{The model}\label{prelim}

Consider the electromagnetic field produced by an electron of mass $m,$ which is on average at rest and interacts with the virtual as well as real photons in equilibrium at finite temperature\footnote{We use relativistic units $\hbar = c = 1.$ Also, the Minkowski metric is $\eta_{\mu\nu} = {\rm diag}\{+1,-1,-1,-1\}.$} $T \ll m.$ The effective electric potential is
\begin{eqnarray}\label{eff}&&
A_0^{\rm eff} (x) = \EuScript{N}^{-1}{\rm Tr} \left( A_0 (x)e^{- \beta H_\phi } \varrho\right)\,, \quad \EuScript{N} = {\rm Tr} \left( e^{- \beta H_\phi } \varrho\right)\,, \quad \beta = 1/T\,.
\end{eqnarray}
\noindent Here $ A_0 (x)$ is the Heisenberg picture operator of the scalar potential, $H_\phi$ the Hamiltonian of free photons, $\varrho$ the electron density matrix, and the trace is over all photon states as well as the single electron states. Having chosen $H_\phi$ to be the free photon Hamiltonian we thereby omit corrections to the photon distribution due to their interactions. This simplification is in fact a valid approximation in investigating the infrared problem, because it is the low-energy asymptotic of the photon distribution that is only important in this case (at the one-loop level, corrections to the photon distribution do not contribute to the effective field at all). Also, the electron Hamiltonian might have been added to the exponent in Eq.~(\ref{eff}). However, since there is only one nonrelativistic electron, and $T\ll m,$ its contribution can be easily shown to factorize and cancel the corresponding contribution to the normalization factor $\EuScript{N},$ leaving the value of the effective field unchanged.

To evaluate the effective field, we use the general framework of the real time approach \cite{realtime} according to which the right hand side of Eq.~(\ref{eff}) can be written in the interaction picture
\begin{eqnarray}\label{effInt}&&
A_0^{\rm eff} (x) = \EuScript{N}^{-1}{\rm Tr} \left( T_c \left[ {\exp i\left( {\int\limits_C {d^4 x} L_I (x)} \right) A_0 (x)} \right] e^{ - \beta  H_\phi  } \varrho\right)\,,
\end{eqnarray}
where $L_I$ is the interaction Lagrangian, the $x^0$-integration is along the standard Schwinger-Keldysh time-contour $C$ running from from $t_i$ to $t_f>x^0$ and back, and $T_c$ denotes the operator ordering along this contour. The conventional limit $t_i \to -\infty$ is directly related to the infrared problem and will be discussed in the last section. As usual, the $T_c$-ordering renders the field propagators $2\times 2$ matrices which in momentum space take the form
\begin{eqnarray}\label{electronpr2}
M(k)\left(\begin{array}{cc}
D_F(k) & d(k_0,\bm{k})\\
d(-k_0,\bm{k}) & -\tilde D_F(k)
\end{array}\right)M(k), \quad M(k) &=& \left(\begin{array}{cc}
1 & \pm\theta(-k_0) \\
\theta(k_0) & 1 \label{M}
\end{array}\right), \nonumber
\end{eqnarray}\noindent where the step function $\theta(-k_0)$ is taken with the plus (minus) sign for the photon (electron), the tilde symbolizes the special operation of complex conjugation with respect to which the Dirac matrices are real, $D_F(k)$ is the usual vacuum Feynman propagator, and $d(k_0,\bm{k}) = - 2\pi i \theta(k_0)n(\bm{k})\delta(k^2).$ In the present case of a single electron system, one has $$D_F(k) =\frac{\slashed{k} + m}{m^2-k^2-i0}\,, \quad n(\bm{k})=0$$ for the electron, while for the photon in the Feynman gauge, $$D_F(k) = \frac{\eta_{\mu\nu}}{k^2 + i0}\,, \quad n(\bm{k})=\frac{1}{e^{\beta|\bm{k}|} - 1}.$$

Upon transition to the momentum space, the effective potential takes the form
\begin{eqnarray}\label{Fur_e}
A^{\rm eff}_{0}(x) = -e\sum\limits_{\sigma ,\sigma '}\iint
\frac{d^3 \bm{q}}{(2\pi)^3} \frac{d^3\bm{p}}{(2\pi)^3}e^{i px}D^{(11)}(p) \varrho_{\sigma \sigma'}(\bm{q},\bm{q} + \bm{p})R(p,q)\bar{u}_{\sigma'}(\bm{q}+\bm{p})\gamma^0u_{\sigma}(\bm{q})\,.
\end{eqnarray}
\noindent In this formula, the values of $q^0$ and $p^0$ are fixed by the mass-shell condition ${q^0 = \sqrt{\bm{q}^2 + m^2}>0}$ and the energy-momentum conservation. Since the field-producing charge is non-relativistic, ${p^0 = (\bm{q}+\bm{p})^2/2m - \bm{q}^2/2m\,.}$ The bispinor amplitudes and the momentum-space density matrix are normalized on unity: $$\bar{u}_{\sigma}u_{\sigma}=1, \quad \sum\limits_{\sigma}\int
\frac{d^3 \bm{q}}{(2\pi)^3}\varrho_{\sigma \sigma}(\bm{q},\bm{q}) = 1\,.$$ The radiative corrections to the effective field are incorporated in the scalar function $R(p,q)$ which is equal to unity in the tree approximation. As is well-known, the notion of one-particle density matrix is of limited validity in relativistic quantum theory because of the possibility of pair creation. However, under the assumption $T\ll m$ the probability of this process is negligible, and the formula (\ref{Fur_e}) shows that the quantity $$\varrho^{\rm eff}_{\sigma \sigma'}(\bm{q},\bm{q}') \equiv \varrho_{\sigma \sigma'}(\bm{q},\bm{q}')R(q'-q,q), \quad q^2 = q'^2 = m^2,$$ is to be considered as an effective density matrix of the electron. If one discards the radiative corrections and the density matrix $\varrho$ describes the state in which the electron is spatially localized near the point $\bm{x}_0$ at time $t,$ then at distances large compared to the characteristic length of the electron spreading, one can write $\varrho_{\sigma \sigma'}(\bm{q},\bm{q} + \bm{p})\approx \varrho_{\sigma \sigma'}(\bm{q},\bm{q})e^{i\bm{p}\bm{x}_0},$ $p^0 \approx 0,$ and also neglect $\bm{p}$ in the bispinor amplitudes. Using the normalization conditions in Eq.~(\ref{Fur_e}) thus yields in this case
\begin{eqnarray}
A^{\rm eff}_{0}(x) = - e \sum\limits_{\sigma ,\sigma '}\iint
\frac{d^3 \bm{q}}{(2\pi)^3} \frac{d^3\bm{p}}{(2\pi)^3}\frac{e^{-i \bm{p}(\bm{x}-\bm{x}_0)}}{-\bm{p}^2} \varrho_{\sigma \sigma'}(\bm{q},\bm{q})\bar{u}_{\sigma'}(\bm{q})\gamma^0u_{\sigma}(\bm{q}) = \frac{e}{4\pi r}\,,\nonumber
\end{eqnarray}
\noindent where $r = |\bm{x}-\bm{x}_0|,$ {\it i.e.,} the Coulomb law. Our aim below is to show that account of the radiative corrections gives $A^{\rm eff}_{0}(x) = 0$ at any finite distance.

\subsection{$\lambda$-regularization}\label{lambdareg}

Infrared singularities in the electromagnetic formfactors and self-energy contributions to the effective field, as well as ultraviolet divergences require intermediate regularization. First of all, we introduce the usual infrared regulator $\lambda_0$ restricting all loop momenta to $k>\lambda_0$ (in the nonzero temperature case, this cutoff is assumed to satisfy $\lambda_0 \ll T$), and also a momentum threshold $\Lambda$ such that $T\ll \Lambda\ll m,$ which identifies the photons with $\lambda_0 < k < \Lambda$ as ``soft.'' As to the usual ultraviolet divergences, they are supposed to be regularized using some conventional means, say, the dimensional technique. The introduced cutoff is still insufficient for the purpose of regularizing diagrams with the self-energy insertions. In order to make these and the corresponding counterterm diagrams meaningful, it is necessary to regularize the mass-shell propagators.\footnote{Leaving the mass shell does not help in this respect, as it precludes factorization of the infrared contributions, to be proved in the next section.} For this purpose, we introduce the following smearing of the Dirac delta-functions expressing conservation of the 4-momentum in the interaction vertices
$\delta^4(q_1+k-q_2) \rightarrow \Delta_\lambda(q_1+k-q_2),$
where $\Delta_\lambda(w)$ satisfies
\begin{eqnarray}
&&\int d^4 w \Delta_\lambda(w)=1, \quad \Delta_\lambda(w)=\Delta_\lambda(-w), \nonumber\\
&&\Delta_\lambda(w\ne 0)  \rightarrow 0 \quad \text{as} \quad \lambda \rightarrow 0.\nonumber
\end{eqnarray}\noindent A convenient choice is
\begin{eqnarray}
\Delta_\lambda(w) = \frac{1}{\pi^2 \lambda^4}\exp\left({-\frac{w_0^2+\bm{w}^2}{\lambda^2}}\right)\,.\nonumber
\end{eqnarray}\noindent It will be assumed in what follows that the parameter $\lambda$ characterizing the width of the smeared delta-function satisfies $\lambda \ll \lambda_0.$

\begin{figure}[h]
\begin{center}
     \includegraphics{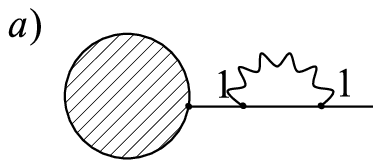}
     \includegraphics{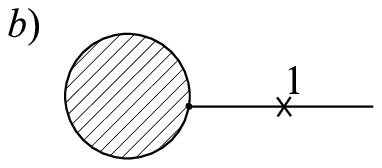}
\caption{(a) Self-energy insertion into external line of a general diagram, (b) the corresponding counterterm diagram.}
\label{CKont}
\end{center}
\end{figure}

Next, to ensure convergence of the effective field in the limit $\lambda\to 0$ (with fixed $\lambda_0\ne 0$), we have to introduce counterterms into Lagrangian. To be specific, let us consider diagrams of the type shown in Fig. \ref{CKont}(a). The external line with the self-energy insertion contributes a factor
\begin{eqnarray}&&\label{factorex}
-\int d^4 w_1 \int d^4 w_2 \Delta_\lambda(w_1)\Delta_\lambda(w_2) \frac{1}{2q(w_1+w_2)+i0}(\slashed{q}+\slashed{w_1}+\slashed{w_2}+m)\Sigma^{(11)}(q + w_1) \nonumber\\&& = - \int d^4 w \Delta^*_\lambda(w) \frac{1}{2qw+i0}(\slashed{q}+\slashed{w}+m)\left[\Sigma^{(11)}(q) + O(\lambda)\right],
\end{eqnarray}\noindent where
$$\Delta^*_\lambda(w) \equiv \frac{1}{16}\int d^4 \xi \Delta_\lambda\left(\frac{w-\xi}{2}\right)\Delta_\lambda\left(\frac{w+\xi}{2}\right).$$
In Eq.~(\ref{factorex}), the electron self-energy $\Sigma^{(11)}(q)$ is taken on the mass shell. The standard renormalization prescription that the propagator poles be at the physical mass requires vanishing of this quantity, which can be met by introducing a counterterm (this prescription is usually realized along with the ultraviolet renormalization). The above expression shows that the delta-functions appearing in the two-point counterterm vertices are to be regularized as
$\delta^4(q_1-q_2) \rightarrow \Delta^*_\lambda(q_1-q_2).$ The counterterm diagram shown in Fig. \ref{CKont}(b) then reads
\begin{eqnarray}\label{contrDelta}
\int d^4 w \Delta^*_\lambda(w) \frac{1}{2qw+i0}(\slashed{q}+\slashed{w}+m)\Sigma^{(11)}(q).
\end{eqnarray}\noindent
The remaining contribution coming from integration of the $O(\lambda)$-term in Eq.~(\ref{factorex}) will be taken care of below when summing the infrared parts of many-loop diagrams. It is interesting to note that in the limit $p \rightarrow 0,$ the counterterm diagrams cancel each other in the expression for the effective field, which can be verified directly using the above definitions. However, this property is violated at finite $p$ by the Lorentz-noninvariant integration in Eq.~(\ref{contrDelta}).

\subsection{Factorization of infrared contributions}\label{factorization}
\begin{figure}[h]
\begin{center}
     \includegraphics{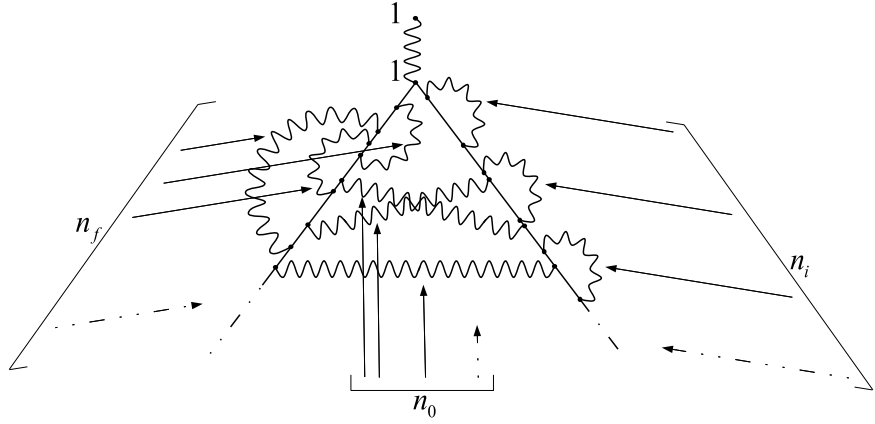}
\caption{General infrared-divergent contribution to the effective field.}
\label{SumA}
\end{center}
\end{figure}

Now that the finite-momenta contributions to the electron self-energy have been canceled by counterterms as discussed in the preceding section, it remains to take into account  contributions of small virtual photon momenta. The general structure of diagrams to be considered is shown in Fig. \ref{SumA}. In terms of the function $R(p,q),$ the sum of all such diagrams can be written as a series

\begin{eqnarray}\label{series}
R(p,q) = I_{\Lambda}(p,q)\sum_{N=0}^{\infty}(e^2)^N I_N (p,q,\Lambda),
\end{eqnarray}\noindent
where the factor $I_{\Lambda}(p,q)$ corresponds to photons with momenta $k > \Lambda.$ $I_N$ has the following general form
\begin{eqnarray}\label{I_N}
I_N (p,q,\Lambda)&&=\frac{1}{i^N}\sum_{n_i=0}^N \sum_{n_f=0}^{N-n_i}\int \frac{d^4 k_1}{(2\pi)^4}\cdots\frac{d^4 k_{n_0}}{(2\pi)^4}
m^{2(n_i+n_f)}(m^2-p^2/2)^{n_0} D^{(11)}(k_1)\cdots D^{(11)}(k_{n_0}) \nonumber \\
&&\times \frac{F_{n_i}^{n_0}(q,k_1,\dots,k_{n_0})F_{n_f}^{n_0}(q+p,k_1,\dots,k_{n_0})}{n_i!n_f!n_0!} + \text{2-terms},
\end{eqnarray}\noindent
where $N$ is the number of virtual photon lines, of which $n_i$ ($n_f$) reside on the ingoing (outgoing) electron line, while the remaining $n_0 \equiv N-n_i-n_f$ connect the two electron lines; ``2-terms'' denotes the contribution of diagrams involving 2-vertices. The function $F_n^{n_0}$ reads
\begin{eqnarray}\label{F_n_Cut}
F_{n}^{n_0}(q,k_1, &&\hspace{-0,3cm} \dots,k_{n_0}) =\int d^4 k_{n_0+1}\cdots d^4 k_{n_0+n} \int d^4 w_1\cdots d^4 w_{2n+n_0}
D^{(11)}(k_{n_0+1})\cdots D^{(11)}(k_{n_0+n}) \nonumber \\
&&\times \Delta(w_{2n+1}-k_1)\cdots \Delta(w_{2n+n_0}-k_{n_0})
\nonumber \\
&& \times \Delta(w_1-k_{n_0+1})\cdots\Delta(w_{2n-1}-k_{n_0+n}) \Delta(w_2+k_{n_0+1})\cdots\Delta(w_{2n}+k_{n_0+n})
\nonumber \\
&& \times \sum\limits_{\rm perm}\left[ \frac{1}{w_1 q+i0} \frac{1}{(w_1+w_2)q+i0}\cdots\frac{1}{(w_1+ \cdots +w_{2n+n_0})q+i0} \right].
\end{eqnarray}\noindent
where the sum is over all permutations of indices $1,2,\dots,2n+n_0.$

\begin{figure}[h]
\begin{center}
     \includegraphics{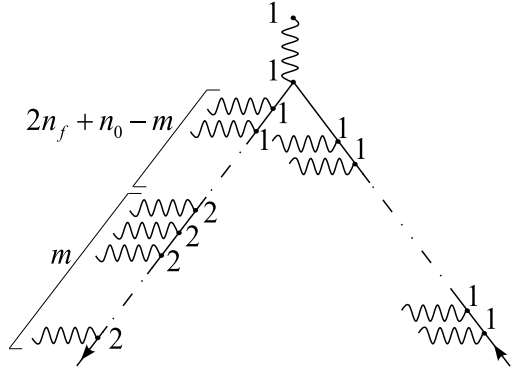}
\caption{Typical diagram involving 2-vertices. It is understood that the horizontal photon lines are arbitrarily paired.}
\label{SumB}
\end{center}
\end{figure}

Let us show that the 2-terms do not contribute to the effective field in the $\lambda$-regularization. Figure \ref{SumB} depicts the general structure of the corresponding diagrams. Graphs with a 1-vertex appearing to the left of a 2-vertex can be omitted, because they involve the function $D_e^{(12)}(q-k)\sim\theta (k_0-q_0),$ and therefore do not contribute at small loop momenta. If a diagram has $m > 1$ vertices of the 2-type on the outgoing electron line, then the corresponding sum over permutations reads
\begin{eqnarray}&&
\hspace{-0,3cm}\sum\limits_{\rm perm}\left[ \frac{1}{w_1 q-i0}\cdots \frac{1}{(w_1 +\cdots +w_{m-1}) q-i0}
\left(\frac{1}{(w_1+\cdots +w_m) q+i0}\right. \right. \nonumber \\
&&\left. \left. \hspace{-0,3cm}-\frac{1}{(w_1 +\cdots +w_m) q-i0}\right) \frac{1}{(w_1 +\cdots +w_{m+1}) q+i0}
\dots \frac{1}{(w_1+\cdots +w_{2n_f+n_0})q+i0}  \right]=\nonumber \\&&
\hspace{-0,3cm}=\sum\limits_{\rm perm}\frac{1}{m!}\left[ \left(\frac{1}{w_1 q-i0}\cdots\frac{1}{w_m q-i0} - \frac{1}{w_1 q-i0}\cdots\frac{1}{w_m q-i0}\right)\frac{1}{(w_1 +\cdots +w_{m+1}) q+i0}\cdots\right]\nonumber\\&& \hspace{-0,3cm} =0,\nonumber
\end{eqnarray}\noindent
where we explicitly performed the permutation over indices $1,...,m.$ In the case $m=1,$ expression in the square brackets takes the form
\begin{eqnarray}
\left(\frac{1}{w_1 q+i0}-\frac{1}{w_1 q-i0}\right)\frac{1}{(w_1+w_2) q+i0}\cdots\frac{1}{(w_1 +\cdots +w_{2n_f+n_0}) q+i0}. \nonumber
\end{eqnarray}\noindent
This vanishes too, because $w_i \neq 0$ in the $\lambda$-regularization, thus proving that the contribution of the 2-terms is zero. Using the Ward identities for the vertex functions, it can be shown that this disappearance of the 2-terms is in fact a non-perturbative result which holds true in other regularizations only if the imaginary part of the electron self-energy vanishes on the mass shell. This is known to be the case at $T=0,$ but not generally at finite temperatures \cite{semland}.

This result permits factorization of the infrared contributions and allows us to sum the series (\ref{series}). Indeed, the sum over permutations in Eq.~(\ref{F_n_Cut}) becomes
\begin{eqnarray}
\frac{1}{w_1 q+i0} \frac{1}{w_2 q+i0}\cdots \frac{1}{w_{2n+n_0}q+i0}\,.\nonumber
\end{eqnarray}\noindent and the series summation can be carried out in a way similar to the standard treatment of the loop infrared divergences (see, {\it e.g.}, \cite{Weinberg}).
Substituting Eq.~(\ref{F_n_Cut}) into Eq.~(\ref{I_N}), and designating
\begin{eqnarray}
d_{st}=-i q_s q_t \int \frac{d^4 k}{(2\pi)^4} \int d^4 w_1 \int d^4 w_2
\frac{\Delta(w_1-\eta_s k)\Delta(w_2+\eta_t k) D^{(11)}(k)}{(w_1 q_s+i0)(w_2 q_t+i0)}, \nonumber
\end{eqnarray}\noindent
where $s,t=1,2$, $q_1=q$, $q_2=q+p$, $\eta_1=1$, $\eta_2=-1$, we obtain
\begin{eqnarray}\label{inpq}
I_N (p,q,\Lambda)=\sum_{n_i=0}^N \sum_{n_f=0}^{N-n_i}
\frac{d_{11}^{n_i}}{n_i!\,2^{n_i}} \frac{d_{22}^{n_f}}{n_f!\,2^{n_f}} \frac{d_{12}^{n_0}}{n_0!}=
\frac{(d_{11}+d_{22}+2d_{12})^N}{N!\,2^N}\,.\nonumber
\end{eqnarray}\noindent
Putting this into Eq.~(\ref{series}) yields
\begin{eqnarray}\label{Jpq}
R(p,q) = \exp\left\{\frac{e^2}{2}(d_{11}+d_{22}+2d_{12})\right\} I_{\Lambda}(p,q) \nonumber
\end{eqnarray}\noindent
In the limit of removed smearing of the $\delta$-functions, $d_{st}$ read
\begin{eqnarray}
d_{st} = i\eta_s \eta_tq_s q_t
\int \frac{d^4 k}{(2\pi)^4} \frac{D^{(11)}(k)}{(k q_s)(k q_t)}\,, \quad \lambda = 0,\nonumber
\end{eqnarray}\noindent
and evaluation of these integrals in the case of small momentum transfer ($|\bm{p}|\ll m$) gives
$$
R(p,q) = \left\{
\begin{array}{lc}
\displaystyle \exp\left(- \frac{\alpha \bm{p}^2}{3\pi m^2}\ln\frac{\Lambda}{\lambda_0} \right)I_{\Lambda}(p,q),& T = 0,\\
\displaystyle\exp\left(- \frac{\alpha \bm{p}^2}{3\pi m^2}\left[\frac{2T}{\lambda_0}+\ln\frac{\Lambda}{T} \right]\right)I_{\Lambda}(p,q),&  T \ne 0\,,
\end{array}
\right.
$$\noindent where $\alpha = e^2/4\pi$ is the fine structure constant.

\subsection{The interpretation}\label{The interpretation}

The above expressions for $R(p,q)$ tell us that the Coulomb field of the electron vanishes at any finite $\bm{x}.$ Indeed, it is seen from Eq.~(\ref{Fur_e}) that this field is determined by the Fourier components with $|\bm{p}|\sim 1/|\bm{x}-\bm{x}_0|,$ while the infrared exponent in the function $R(p,q)$ tends to zero for $\bm{p}\ne 0$ in the limit $\lambda_0 \to 0.$ On the other hand, this exponent equals unity for $\bm{p} = 0,$ which expresses the electric charge conservation. In fact, it is not difficult to verify validity of the Gauss law in  infinite space: In $\bm{p}$-representation, integrating the electric field over an infinitely remote sphere is equivalent to multiplying the integrand of Eq.~(\ref{Fur_e}) by $\bm{p}^2$ and taking the limit $\bm{p}\to 0,$
$$\lim\limits_{\bm{p}\to 0}\sum\limits_{\sigma ,\sigma '}\int
\frac{d^3 \bm{q}}{(2\pi)^3}\bm{p}^2(-e)D^{(11)}(p) \varrho_{\sigma \sigma'}(\bm{q},\bm{q} + \bm{p})R(p,q)u^*_{\sigma'}(\bm{q}+\bm{p})u_{\sigma}(\bm{q}) = e\sum\limits_{\sigma}\int\frac{d^3 \bm{q}}{(2\pi)^3}\varrho_{\sigma \sigma}(\bm{q},\bm{q}),$$ which is equal to the electron charge by virtue of the normalization condition for $\varrho.$

The natural physical interpretation of this result is that the interaction with soft photons causes an electron to spread over infinite space so that the charge density becomes infinitely small everywhere. An essential difference of this spreading from the usual one is that it takes place independently of the particular form of the electron density matrix, while in nonrelativistic quantum mechanics, the density matrix of a free electron can always be chosen so as to describe a state which is spatially localized at any given time instant. As was already mentioned, the expression (\ref{Fur_e}) for the effective electric field suggests that the matrix $\varrho_{\sigma \sigma'}(\bm{q},\bm{q}')R(q'-q,q)$ is to be considered as an effective density matrix of the electron, which incorporates the effects of its interaction with soft photons. In terms of this matrix, the electron spreading can be described as its thermalization. Indeed, the fact that the function $R(q'-q,q)$ vanishes for $\bm{q}\ne \bm{q}'$ in the limit $\lambda_0 \to 0$ means that the effective density matrix becomes diagonal, and hence time-independent in this limit, signifying that the electron is driven near an equilibrium with photons.

It is remarkable that this thermalization occurs even at zero temperature, the difference from the case $T \ne 0$ being only quantitative: the power dependence on the infrared cutoff at $T=0$ switches to an exponential dependence at finite temperatures. Yet, this difference is important from the practical point of view.
To assess the influence of the infrared exponent, we note that any actual measurement naturally sets an infrared cutoff specific to the given experimental situation. In particular, the total duration of the experiment, $\tau,$ cuts off the photon energy at $\sim\hbar/\tau.$ This implies that instead of taking the formal limit $t_i \to -\infty,$ Eq.~(\ref{effInt}) is to be considered at finite $t_i$ such that $t - t_i \sim \tau,$ which regularizes the energy-integrations in the Feynman integrals. Furthermore, the finite fundamental speed of interaction propagation effectively confines the system to a box with the linear dimension $c\tau,$ thereby cutting off all momenta at $\lambda_0 \sim \hbar/c\tau.$ Noting also that any field measurement is meaningful only at distances exceeding the Compton length, $r \gtrsim l_c=\hbar/mc,$ and replacing the threshold $\Lambda$ by $mc,$ we see that in the zero-temperature case, the infrared exponent becomes important when
$$\frac{\alpha}{3\pi}\ln\frac{mc^2\tau}{\hbar} \sim 1.$$ The corresponding time $\tau \sim 10^{540}$\,s far exceeds the age of the Universe. However, things change at finite temperatures. In experiments using cathode-ray tubes, for instance, $r$ ranges approximately $1\mu$m to 1cm, and the corresponding $$\tau \sim \frac{3\pi\hbar}{2\alpha T}\left(\frac{r}{l_c}\right)^2$$ ranges $10^2$ to $10^{10}$ seconds, at room temperature.

Thus, sufficiently slow experiments involving free electrons may require taking into account the electron thermalization. Still, this effect is completely negligible as far as one is concerned with the electron electric field itself, because of the usual non-relativistic spreading. Indeed, let $a$ denote characteristic length of the electron wavefunction before the field measurement. The subsequent free electron evolution according to the Schr${\rm \ddot{o}}$dinger equation leads to the wavefunction spreading, the characteristic time being $ma^2/\hbar$ ({\it i.e.,} $a^2$ grows with time approximately as $\hbar t/m$). Since a meaningful measurement in any case requires $a\lesssim r,$ it follows from the above expressions that the electron thermalization might affect its field during the experiment only under the condition $$\frac{\hbar}{\alpha T}\left(\frac{r}{l_c}\right)^2 \lesssim \frac{mr^2}{\hbar}\,,$$ or $T\gtrsim mc^2/\alpha.$ But the latter is opposite to the general assumption $T \ll mc^2$ underlying our consideration of the single electron picture.

We arrive at the conclusion that the infrared singularity in the electromagnetic field produced by a free electron is negligible in the description of processes driven by the Coulomb interaction. In particular, it follows from the above discussion that the electron thermalization can be completely discarded in the formulation of the asymptotic conditions for the scattering experiments involving charged particles. However, since this thermalization modifies the electron density matrix, it can in principle affect sufficiently slow processes, in particular, those sensitive to changes in the quantum entropy of electron states.

Further details of the calculation including comparison of different regularization schemes can be found in Ref.~\cite{qed1}.

\acknowledgements{We thank Drs. P.~Pronin, G.~Sardanashvili, and especially A.~Baurov (Moscow State University) for their continuing interest to our work and useful discussions.}

\end{document}